# Dynamical symmetry breaking in quasistatic magnetic oscillations

E.O. Kamenetskii

Department of Electrical and Computer Engineering,
Ben Gurion University of the Negev, Beer Sheva, 84105, ISRAEL

**Abstract**

Recent microwave experiments demonstrate the anapole-moment and magnetoelectric properties in quasi-2D ferrite particles with magnetic-dipolar-wave oscillating spectra. The theory developed in this paper shows that there are the macroscopically quantum topological effects. Quantum coherence for macroscopic systems refers to circumstances when large numbers of particles can collectively cooperate in a single quantum state. These effects are rarely observed through macroscopic measurements because statistical averaging over many states usually masks all evidence of quantum discreteness. Magnetic-dipolar oscillating modes in normally magnetized ferrite disks demonstrate properties of a Hamiltonian system. The purpose of this paper is to show that because of the adiabatic motion process for such a Hamiltonian system one has macroscopic quantum effects of symmetry breaking, magnetic currents, and eigen electric moments.

PACS: 76.50.+g; 68.65.-k; 11.30.Er

# 1. Introduction

In quasi-2D systems, the dipolar interaction can play an essential role in determine the magnetic properties. In these systems the short-range exchange interactions alone are not necessarily sufficient to establish a ferromagnetically ordered ground state. The dipolar interaction is important in stabilizing long-range magnetic order in 2D systems, as well in determining the nature and morphology of the ordered states [1].

In the analysis of the spin-wave spectra in quasi-2D spin systems, including the dipolar interaction considerably complicates the problem. There has been developed the microscopic formalism for the dipole interactions in spin-wave ferromagnetic films [2]. On the other hand, there has been extensive work to generalize the magnetostatic-wave (or continuous-medium) theory to include the exchange effects in two-dimensional ferrites [3]. These complicated dipole-exchange theories should be applicable, however, for ultrathin ferrite films. In microwave experiments on the exchange effects, there are the films with thickness about units of micrometers and less. For films with thickness about tens of micrometers one should use the continuum approach. The continuum theory by its very nature is long wavelength to the extent that it must consider disturbances long compared to the lattice spacing. The continuum analog of the equation of motion of a single spin is the equation of motion of the magnetization vector (the Landau-Lifshitz equation). So one can describe the magnetization dynamics with use of the susceptibility and permeability tensors and solve the problem based on the magnetostatic-wave (MS-wave) theory [4]. As a characteristic feature of the magnetic-dipolar mode spectra in ferrite samples is the fact that for such wave processes there is negligibly small variation of the exchange energy and, at the same time, negligibly small variation of the electric energy.

The theory of the magnetic-dipolar mode spectra is based on the notion of magnetostatic-potential wave functions $\psi$. These functions appear through representation of the RF magnetic fields in a ferrite sample as $\vec{H} = -\nabla \psi$ and usually are considered just as formal quantities convenient for calculations [4]. Really, from the classical electrodynamics it follows that just only the $\vec{E}$ and $\vec{H}$ fields, but not potentials, are regarded as the basic physical quantities. Nevertheless, the fact that in the MS-wave processes one has negligibly small variation of the electric energy raises the questions about the nature of the RF fields in magnetic-dipole oscillations. In particular, the question about the electromagnetic power flow (the Poynting vector) for MS-wave modes arises: There are no physical mechanisms describing the effect of transformation of the curl electric field to the magnetostatic-potential magnetic field.

MS-potential wave functions may acquire, however, a special physical meaning. This is clearly demonstrated in a case of magnetic-dipolar oscillations in a normally magnetized ferrite disk. As it has been shown recently [5-7], in such a sample the confinement effect leads to the quantum-like properties (charcterized by the Schrödinger-like equation) for magnetic-dipolar oscillations, which are completely described by the MS-potential wave functions. The oscillations can be considered as the motion process of certain quasiparticles – the light magnons – having quantization of energy and characterizing by effective masses depending on the energy levels [8, 9]. To a certain extent, one has situation resembling the dipole-interaction "flat" quasiparticles (electron-hole pairs – excitons) in disk-form semiconductor dots [10]. Together with such similarity with semiconductor quantum dots as discrete energy levels due to confinement phenomena, ferrite particles show other, very unique, properties attributed to the quantized-like systems. There are special symmetry properties of the anapole moment [polar (electric) symmetry, i.e. the parity-odd, and time-reversal-even symmetry] known from elementary particle physics [11] and symmetry properties of microwave oscillation spectra observed experimentally in ferrite magnetoelectric (ME) particles [12].

At present, magnetoelectricity is considered as a phenomenon with local coexistence of electric and magnetic moments. In crystals and molecular systems, this phenomenon takes place when space inversion is locally broken [13]. Magnetoelectric interactions with mutually perpendicular electric and magnetic dipoles in crystal structures arise from toroidal distributions of currents and are described by anapole moments [14]. It has been discussed that the reason why such anapole moments appear is Stone's spinning-solenoid Hamiltonian [15]. Magnetoelectric properties of Stone's Hamiltonian become apparent because of Berry's curvature of the electronic wavefunctions. In recent theories of spin waves in magnetic-order crystals, a Berry curvature is stated as playing a key role [16, 17]. The input for the spin-wave equations are the energies and Berry curvatures of many-electron states describing frozen spin spirals. One switches from the idealized Heisenberg model to a real crystal, where the magnetization is a continuum vector. The Berry phase may also influence the properties of magnons. If the magnetic medium in which the magnon is propagating is spatially non uniform, a Berry phase may be accumulated along a closed circuit in space. It has been recently indicated [18] that the geometric Berry phase due to a non-coplanar texture of the magnetization of a ferromagnetic ring would affect the dispersion of magnons, lifting the degeneracy of clockwise and anticlockwise propagating magnons. As it is discussed in [18], in ferromagnetic metals the magnetization current endows the ring with an electric dipole moment.

Quantum effects are rarely observed through macroscopic measurements because statistical averaging over many states usually masks all evidence of discreteness. Notable exceptions include the dc and ac Josephson effects, the dc Haas-van Alphen effect and the quantum Hall effect. Magnetic-dipolar oscillating modes in normally magnetized ferrite disks demonstrate properties of a Hamiltonian system. The purpose of this paper is to show that for magnetic-dipolar oscillating modes in normally magnetized ferrite disks one has such macroscopic quantum effects as symmetry breaking, magnetic currents, and eigen electric moments. The confinement effect for magnetic-dipolar oscillations requires proper phase relationships to guarantee single-valuedness of the wave functions. To compensate for sign ambiguities and thus to make wave functions single valued we add a vector-potential-type term to the MS-potential Hamiltonian. Previously, we stated [11] that an anapole moment of MS oscillations appears due to the handedness properties of the boundary conditions in a normally magnetized ferrite disk. In our present analysis we will use the Waldron's helical coordinate system. Waldron showed [19] that the solution of the Helmholtz equation in a helical coordinate system can be reduced to the solution of the Bessel equation. With use of the Waldron coordinate system, Overfelt had got analytic exact solutions of the Laplace equation in a helical coordinate system with a reference to the helical Bessel functions and helical harmonics for static fields [20]. In this paper we show that a Berry's phase curvature for magnetic-dipolar modes appears explicitly due to the LL motion equation taking into account handedness properties.

## 2. Magnetostatic modes in a helical coordinate system and symmetry breaking

To analyze symmetry properties of magnetic-dipolar spin modes we consider the MS-wave propagation in a helical coordinate system. Such an analysis in an infinite MS-wave waveguide bears a formal character, but acquires real physical meaning in a case of restricted waveguide sections. Let us consider the right-handed and left-handed helical coordinate systems. We make a supposition that for magnetic-dipolar modes in helical coordinates diagonal and off-diagonal components of the permeability tensor remain the same as in the cylindrical coordinate system. This supposition is clearly acceptable since the long-range magnetic-dipolar field variations have no influence on the character of local spin precession. This is especially correct in a case of our geometry of a normally magnetized thin-film ferrite disk. At the same time, for solutions of the Landau-Lifshitz (LL) equation one has different signs of the off-diagonal components of the permeability tensor in the right-handed and left-handed coordinate systems. Lifting the degeneracy of clockwise and anticlockwise propagating MS modes becomes apparent when one considers the LL equation for harmonic RF magnetization: $i\omega \vec{m} + \gamma \vec{m} \times \vec{H}_0 = -\gamma \vec{M}_0 \times \vec{H}$ with respect to a helical coordinate system. For given directions of bias magnetic field $\vec{H}_0$ and saturation magnetization $\vec{M}_0$, and for given rf magnetic field $\vec{H}$ and RF magnetization $\vec{m}$ one has opposite signs for vector products with respect to the right-handed and left-handed helical coordinate systems. This shows that for solutions of the LL equation: $\vec{m} = \vec{\vec{\chi}} \cdot \vec{H}$ there are different signs of off-diagonal components of the susceptibility tensor $\vec{\vec{\chi}}$ with respect to the right-handed and the left-handed helical coordinate systems. For a DC magnetic field directed along $z$-axis one has for the permeability tensor:

$$\vec{\vec{\mu}} = \begin{bmatrix} \mu & \pm i\mu_a & 0 \\ \mp i\mu_a & \mu & 0 \\ 0 & 0 & 1 \end{bmatrix}, \tag{1}$$

where the upper signs correspond to the right-hand helical coordinate system and the lower signs correspond to the left-hand helical coordinate system.

For MS modes in an infinite axially magnetized ferrite rod, components of the magnetic flux density are expressed by means of components of the magnetic field in the Waldron's helical coordinate system $(r, \phi, \zeta)$ [19] as:

$$\begin{aligned} B_r &= \pm i\mu_a H_\phi (\cos\alpha_0)^{(R,L)} + \mu H_r, \\ B_\phi &= \mu H_\phi \mp i \frac{\mu_a}{(\cos\alpha_0)^{(R,L)}} H_r, \\ B_\zeta &= H_\zeta + (1-\mu) H_\phi (\sin\alpha_0)^{(R,L)} + i\mu_a H_r (\tan\alpha_0)^{(R,L)}, \end{aligned} \tag{2}$$

where subscripts $(R,L)$ mean, respectively, right-handed and left-handed helical coordinate systems. The terms corresponding to the right-handed and left-handed coordinates may have different signs. With representation of magnetic field as $\vec{H} = -\nabla\psi$ and with use of transformations in helical coordinates [19], we rewrite Eqs. (2) as:

$$\begin{aligned} B_r &= -\left[\mu \frac{\partial\psi}{\partial r} \pm i\mu_a \left(\frac{1}{r}\frac{\partial\psi}{\partial\phi} - (\tan\alpha_0)^{(R,L)} \frac{\partial\psi}{\partial\zeta}\right)\right], \\ B_\phi &= -\frac{1}{(\cos\alpha_0)^{(R,L)}} \left[\mu\left(\frac{1}{r}\frac{\partial\psi}{\partial\phi} - (\tan\alpha_0)^{(R,L)} \frac{\partial\psi}{\partial\zeta}\right) \mp i\mu_a \frac{\partial\psi}{\partial r}\right], \\ B_\zeta &= -(\tan\alpha_0)^{(R,L)} \left[\frac{2}{(\sin 2\alpha_0)^{(R,L)}} \frac{\partial\psi}{\partial\zeta} - \frac{1}{r}\frac{\partial\psi}{\partial\phi} + (1-\mu)\left(\frac{1}{r}\frac{\partial\psi}{\partial\phi} - (\tan\alpha_0)^{(R,L)} \frac{\partial\psi}{\partial\zeta}\right) \pm i\mu_a \frac{\partial\psi}{\partial r}\right]. \end{aligned} \tag{3}$$

Based on Waldron's equation for the divergence [19], we have

$$\nabla \cdot \vec{B} = \frac{1}{r}\frac{\partial}{\partial r}(rB_r) + \frac{\cos\alpha_0}{r}\frac{\partial B_\phi}{\partial\phi} + \frac{\partial B_\zeta}{\partial\zeta} = 0. \tag{4}$$

After some transformations we obtain the Walker equation in helical coordinates:

$$\frac{\partial^2 \psi}{\partial r^2} + \frac{1}{r}\frac{\partial \psi}{\partial r} + \frac{1}{r^2}\frac{\partial^2 \psi}{\partial \phi^2} + \left(\frac{1}{\mu} + \tan^2 \alpha_0\right)\frac{\partial^2 \psi}{\partial \zeta^2} - 2\frac{1}{r}(\tan \alpha_0)^{(R.L)}\frac{\partial^2 \psi}{\partial \phi \partial \zeta} = 0. \tag{5}$$

Outside a ferrite region (where $\mu = 1$) Eq. (5) reduces to the Laplace equation in helical coordinates [19, 20]. Following Overfelt's approach [20], we assume that solutions of the Laplace and Walker equations are found as

$$\psi(r,\phi,\zeta) = R(r)P(\phi)Z(\zeta), \tag{6}$$

where

$$\begin{aligned} P(\phi) &\sim \exp(\pm i w \phi), \\ Z(\zeta) &\sim \exp(\pm i \beta \zeta). \end{aligned} \tag{7}$$

Here the quantities $w$ and $\beta$ are assumed to be real and positive.

Inside and outside a ferrite rod one has the following four solutions for the MS-potential wave function:

$$\begin{aligned} \psi^{(1)} &\sim e^{-iw\phi}e^{-i\beta\zeta}, \\ \psi^{(2)} &\sim e^{+iw\phi}e^{-i\beta\zeta}, \\ \psi^{(3)} &\sim e^{+iw\phi}e^{+i\beta\zeta}, \\ \psi^{(4)} &\sim e^{-iw\phi}e^{+i\beta\zeta}. \end{aligned} \tag{8}$$

We consider the $\psi^{(1)}$ wave as the forward (propagating in a ferrite rod along $+z$ axis) right-hand-helix (FR) MS wave. For this wave we will take $(\tan \alpha_0)^{(R)} \equiv \tan \alpha_0 \equiv \bar{p}/r$, where $\bar{p} = p/2\pi$, $p$ is the helical pitch. The quantities $\tan \alpha_0$ and $\bar{p}$ are assumed to be positive.

To have the Walker equation (5) for other types of helical waves $(\psi^{(2)}, \psi^{(3)}, \psi^{(4)})$ the same as for wave $\psi^{(1)}$, we should come to conclusion that the wave $\psi^{(2)}$ is the forward left-hand (FL) wave, the wave $\psi^{(3)}$ is the backward right-hand (BR) wave, and the wave $\psi^{(4)}$ is the backward left-hand (BL) wave. We can sum up the above as follows. For the FR wave ($\psi^{(1)}$) and BR wave ($\psi^{(3)}$), there is $(\tan \alpha_0)^{(R)} = \tan \alpha_0 = \bar{p}/r$, and for the FL wave ($\psi^{(2)}$) and BL wave ($\psi^{(4)}$) there is $(\tan \alpha_0)^{(L)} = -\tan \alpha_0 = -\bar{p}/r$.

Functions $\psi(r)$ are described by the Bessel equations. For a ferrite rod with radius $\Re$, we have

$$\frac{\partial^2 \psi(r)}{\partial r^2} + \frac{1}{r}\frac{\partial \psi(r)}{\partial r} - \left[\frac{\beta^2}{\mu} + \frac{1}{r^2}(w - \bar{p}\beta)^2\right]\psi(r) = 0 \tag{9}$$

inside a ferrite rod $(r \leq \Re)$ and

$$\frac{\partial^2 \psi(r)}{\partial r^2} + \frac{1}{r}\frac{\partial \psi(r)}{\partial r} - \left[\beta^2 + \frac{1}{r^2}(w - \bar{p}\beta)^2\right]\psi(r) = 0 \qquad (10)$$

outside a ferrite rod $(r \geq \Re)$. Physically acceptable solutions for Eqs. (9) and (10) are possible only for negative quantities $\mu$. Inside a ferrite region ($r \leq \Re$) the solutions are:

$$\begin{aligned}
\psi^{(1)} &= a_1 J_{(w-\bar{p}\beta)}\left[(-\mu)^{1/2}|\beta|r\right]e^{-iw\phi}e^{-i\beta\zeta}, \\
\psi^{(2)} &= a_2 J_{(w-\bar{p}\beta)}\left[(-\mu)^{1/2}|\beta|r\right]e^{+iw\phi}e^{-i\beta\zeta}, \\
\psi^{(3)} &= a_3 J_{(w-\bar{p}\beta)}\left[(-\mu)^{1/2}|\beta|r\right]e^{+iw\phi}e^{+i\beta\zeta}, \\
\psi^{(4)} &= a_4 J_{(w-\bar{p}\beta)}\left[(-\mu)^{1/2}|\beta|r\right]e^{-iw\phi}e^{+i\beta\zeta}.
\end{aligned} \qquad (11)$$

For an outside region ($r \geq \Re$) one has:

$$\begin{aligned}
\psi^{(1)} &= b_1 K_{(w-\bar{p}\beta)}\left(|\beta|r\right)e^{-iw\phi}e^{-i\beta\zeta}, \\
\psi^{(2)} &= b_2 K_{(w-\bar{p}\beta)}\left(|\beta|r\right)e^{+iw\phi}e^{-i\beta\zeta}, \\
\psi^{(3)} &= b_3 K_{(w-\bar{p}\beta)}\left(|\beta|r\right)e^{+iw\phi}e^{+i\beta\zeta}, \\
\psi^{(4)} &= b_4 K_{(w-\bar{p}\beta)}\left(|\beta|r\right)e^{-iw\phi}e^{+i\beta\zeta}.
\end{aligned} \qquad (12)$$

Here $J$ and $K$ are Bessel functions of real and imaginary arguments, respectively. Coefficients $a_{1,2,3,4}$ and $b_{1,2,3,4}$ are amplitude coefficients.

Now we can obtain proper equations for magnetic flux density components of helical waves. For the FR and FL waves we have, respectively

$$\begin{aligned}
B_r^{(1)} &= -\left[\mu\frac{\partial \psi}{\partial r} + i\mu_a\left(\frac{1}{r}\frac{\partial \psi}{\partial \phi} - \tan\alpha_0 \frac{\partial \psi}{\partial \zeta}\right)\right], \\
B_\phi^{(1)} &= -\frac{1}{\cos\alpha_0}\left[\mu\left(\frac{1}{r}\frac{\partial \psi}{\partial \phi} - \tan\alpha_0 \frac{\partial \psi}{\partial \zeta}\right) - i\mu_a\frac{\partial \psi}{\partial r}\right], \\
B_\zeta^{(1)} &= -\tan\alpha_0\left[\frac{2}{\sin 2\alpha_0}\frac{\partial \psi}{\partial \zeta} - \frac{1}{r}\frac{\partial \psi}{\partial \phi} + (1-\mu)\left(\frac{1}{r}\frac{\partial \psi}{\partial \phi} - \tan\alpha_0 \frac{\partial \psi}{\partial \zeta}\right) + i\mu_a\frac{\partial \psi}{\partial r}\right].
\end{aligned} \qquad (13)$$

and

$$\begin{aligned}
B_r^{(2)} &= -\left[\mu\frac{\partial \psi}{\partial r} - i\mu_a\left(\frac{1}{r}\frac{\partial \psi}{\partial \phi} + \tan\alpha_0 \frac{\partial \psi}{\partial \zeta}\right)\right], \\
B_\phi^{(2)} &= -\frac{1}{\cos\alpha_0}\left[\mu\left(\frac{1}{r}\frac{\partial \psi}{\partial \phi} + \tan\alpha_0 \frac{\partial \psi}{\partial \zeta}\right) + i\mu_a\frac{\partial \psi}{\partial r}\right], \\
B_\zeta^{(2)} &= \tan\alpha_0\left[-\frac{2}{\sin 2\alpha_0}\frac{\partial \psi}{\partial \zeta} - \frac{1}{r}\frac{\partial \psi}{\partial \phi} + (1-\mu)\left(\frac{1}{r}\frac{\partial \psi}{\partial \phi} + \tan\alpha_0 \frac{\partial \psi}{\partial \zeta}\right) - i\mu_a\frac{\partial \psi}{\partial r}\right].
\end{aligned} \qquad (14)$$

Now let us consider the BR wave. This wave propagates in the right-hand coordinate system but along $-z$ axis. So compared to the equations (13) for the FR wave, there should be opposite signs before the terms containing quantity $\mu_a$. We have for the BR wave

$$B_r^{(3)} = -\left[\mu\frac{\partial\psi}{\partial r} - i\mu_a\left(\frac{1}{r}\frac{\partial\psi}{\partial\phi} - \tan\alpha_0\frac{\partial\psi}{\partial\zeta}\right)\right],$$

$$B_\phi^{(3)} = -\frac{1}{\cos\alpha_0}\left[\mu\left(\frac{1}{r}\frac{\partial\psi}{\partial\phi} - \tan\alpha_0\frac{\partial\psi}{\partial\zeta}\right) + i\mu_a\frac{\partial\psi}{\partial r}\right], \quad (15)$$

$$B_\zeta^{(3)} = -\tan\alpha_0\left[\frac{2}{\sin 2\alpha_0}\frac{\partial\psi}{\partial\zeta} - \frac{1}{r}\frac{\partial\psi}{\partial\phi} + (1-\mu)\left(\frac{1}{r}\frac{\partial\psi}{\partial\phi} - \tan\alpha_0\frac{\partial\psi}{\partial\zeta}\right) - i\mu_a\frac{\partial\psi}{\partial r}\right].$$

Similarly, changing signs in Eqs. (14) before the terms containing quantity $\mu_a$, we have for the BL wave:

$$B_r^{(4)} = -\left[\mu\frac{\partial\psi}{\partial r} + i\mu_a\left(\frac{1}{r}\frac{\partial\psi}{\partial\phi} + \tan\alpha_0\frac{\partial\psi}{\partial\zeta}\right)\right],$$

$$B_\phi^{(4)} = -\frac{1}{\cos\alpha_0}\left[\mu\left(\frac{1}{r}\frac{\partial\psi}{\partial\phi} + \tan\alpha_0\frac{\partial\psi}{\partial\zeta}\right) - i\mu_a\frac{\partial\psi}{\partial r}\right], \quad (16)$$

$$B_\zeta^{(4)} = \tan\alpha_0\left[-\frac{2}{\sin 2\alpha_0}\frac{\partial\psi}{\partial\zeta} - \frac{1}{r}\frac{\partial\psi}{\partial\phi} + (1-\mu)\left(\frac{1}{r}\frac{\partial\psi}{\partial\phi} + \tan\alpha_0\frac{\partial\psi}{\partial\zeta}\right) + i\mu_a\frac{\partial\psi}{\partial r}\right].$$

The above four helical waves should be considered as components of MS-potential function and space components of magnetic flux density. So we have the following four-component functions:

$$[\psi] = \begin{pmatrix}\psi^{(1)}\\\psi^{(2)}\\\psi^{(3)}\\\psi^{(4)}\end{pmatrix}, \quad [B_r] = \begin{pmatrix}B_r^{(1)}\\B_r^{(2)}\\B_r^{(3)}\\B_r^{(4)}\end{pmatrix}, \quad [B_\phi] = \begin{pmatrix}B_\phi^{(1)}\\B_\phi^{(2)}\\B_\phi^{(3)}\\B_\phi^{(4)}\end{pmatrix}, \quad [B_\zeta] = \begin{pmatrix}B_\zeta^{(1)}\\B_\zeta^{(2)}\\B_\zeta^{(3)}\\B_\zeta^{(4)}\end{pmatrix}. \quad (17)$$

On a cylindrical surface of a ferrite rod we have the boundary conditions:

$$(\psi)_{r=\Re^-} = (\psi)_{r=\Re^+} \quad \text{and} \quad (B_r)_{r=\Re^-} = (B_r)_{r=\Re^+}. \quad (18)$$

In a general form, the boundary condition for radial components of the magnetic flux density [see Eqs. (13) – (16)] can be rewritten as

$$\left[\mu\left(\frac{\partial\psi}{\partial r}\right) \pm i\mu_a\frac{1}{\Re}\left(\frac{\partial\psi}{\partial\phi} \mp \overline{p}\frac{\partial\psi}{\partial\zeta}\right)\right]_{r=\Re^-} = \left(\frac{\partial\psi}{\partial r}\right)_{r=\Re^+}. \quad (19)$$

Based on the above Bessel equations and boundary conditions one obtains a characteristic equation for helical MS waves in a ferrite rod

$$(-\mu)^{1/2} \frac{J'_{(w-\bar{p}\beta)}}{J_{(w-\bar{p}\beta)}} + \frac{K'_{(w-\bar{p}\beta)}}{K_{(w-\bar{p}\beta)}} + \frac{\mu_a (w-\bar{p}\beta)}{|\beta|\Re} = 0, \tag{20}$$

where the prime denotes differentiation with respect to the argument. Eq. (20) is relevant for all helical modes: $\psi^{(1)}$, $\psi^{(2)}$, $\psi^{(3)}$, and $\psi^{(4)}$.

For a smooth infinite ferrite rod, the four-type helical waves (the forward and backward waves with right-handed and left-handed rotations) are basic solutions which cannot be mutually transformed [20]. So it looks that the above analysis bears a formal character and does not have any physical meaning. Situation, however, becomes different in a case of a ferrite disk. In a disk the quantity of pitch $\bar{p}$ can be determined by the virtual "reflection" planes. The distance between virtual "reflection" planes can be considered as an effective disk thickness $d^{eff}$. One can take pitch $p$ equal to $d^{eff}$.

One can see from Eqs. (13) and (16) that for helical waves $\psi^{(1)}$ and $\psi^{(4)}$ the terms containing $\mu_a$ have the same signs. It means that these waves (propagating in opposite directions of $z$-axis but at the same direction of azimuth rotation) are mutually reciprocal. Similar correlation for helical waves $\psi^{(2)}$ and $\psi^{(3)}$ one has from Eqs. (14) and (15). At the same time, waves $\psi^{(1)}$ and $\psi^{(3)}$ as well as waves $\psi^{(2)}$ and $\psi^{(4)}$ are not mutually reciprocal. For mutually reciprocal waves one can consider the closed-loop way. For waves $\psi^{(1)}$ and $\psi^{(4)}$, the closed-loop way has two parts: the forward part of the way is along the FR wave $\psi^{(1)}$ and the backward part of the way – along the BL wave $\psi^{(4)}$. To have such a closed-loop way, one should assume that

$$w^{(1)} = w^{(4)} \equiv w^{(1,4)}, \quad \beta^{(1)} = \beta^{(4)} \equiv \beta^{(1,4)}, \quad \text{and} \quad \bar{p}^{(1)} = \bar{p}^{(4)} \equiv \bar{p}^{(1,4)}. \tag{21}$$

Similarly, we have the closed-loop way for waves $\psi^{(2)}$ and $\psi^{(3)}$ if

$$w^{(2)} = w^{(3)} \equiv w^{(2,3)}, \quad \beta^{(2)} = \beta^{(3)} \equiv \beta^{(2,3)}, \quad \text{and} \quad \bar{p}^{(2)} = \bar{p}^{(3)} \equiv \bar{p}^{(2,3)}. \tag{22}$$

Eqs. (21) and (22) describe the resonance conditions for a ferrite disk resonator which is considered as a section of a ferrite rod waveguide. The first resonance is due to $\psi^{(1)} \leftrightarrow \psi^{(4)}$ interaction. The second resonance is due to $\psi^{(2)} \leftrightarrow \psi^{(3)}$ interaction. These resonances, being characterized by different directions of azimuth rotation, we will distinguish by subscripts, respectively, (+) and (–). We introduce now the following quantities:

$$v_{(+)} \equiv w^{(1,4)} - \bar{p}^{(1,4)} \beta^{(1,4)} \tag{23}$$

and

$$v_{(-)} \equiv w^{(2,3)} - \bar{p}^{(2,3)} \beta^{(2,3)}. \tag{24}$$

Based on Eq. (20) we have for two resonances:

$$(-\mu)^{1/2} \left( \frac{J'_{v_{(+)}}}{J_{v_{(+)}}} \right)_{r=\Re} + \left( \frac{K'_{v_{(+)}}}{K_{v_{(+)}}} \right)_{r=\Re} + \frac{\mu_a v_{(+)}}{|\beta^{(1,4)}|\Re} = 0 \tag{25}$$

and

$$(-\mu)^{1/2}\left(\frac{J'_{\nu_{(-)}}}{J_{\nu_{(-)}}}\right)_{r=\Re} + \left(\frac{K'_{\nu_{(-)}}}{K_{\nu_{(-)}}}\right)_{r=\Re} + \frac{\mu_a \nu_{(-)}}{\left|\beta^{(2,3)}\right|\Re} = 0. \tag{26}$$

On virtual perfect-reflection planes, reflections of helical MS waves take place. For these helical waves one has the failure of the law of reflection symmetry. It means that any helical wave "incident" on a virtual reflection plane cannot be transformed to itself and should be transformed to another-type helical wave. In other words, on a reflection plane one has coupling between different helical waves (between waves with different types of symmetry). The effect of transformations of different types of helical modes in the virtual reflection points give examples of nontrivial bundles for one-dimensional arrays [15].

To have mutual transformations of helical modes stipulating the (+) and (−) resonances ($\psi^{(1)} \leftrightarrow \psi^{(4)}$ and $\psi^{(2)} \leftrightarrow \psi^{(3)}$), there should be certain phase correlations for MS-potential wave functions and components of the magnetic flux density in the reflection points. From Eqs. (13) – (16) one can see that the following equations take place for the resonant modes:

$$B_r^{(1)} = B_r^{(4)}, \; B_\theta^{(1)} = B_\theta^{(4)} \tag{27}$$

and

$$B_r^{(2)} = B_r^{(3)}, \; B_\theta^{(2)} = B_\theta^{(3)}, \tag{28}$$

where $B_\theta = B_\phi \cos\alpha_0$ is the azimuth component of the magnetic flux density in cylindrical coordinates ($r, \theta, z$) [19]. These equations should be taken into consideration to analyze the virtual effect of transformation of one type of a helical mode to another type in the reflection points.

Since $\beta^{(1,4)} \neq \beta^{(2,3)}$, to get the (+) and (−) resonances one should use virtual-reflection-plane disks with different thicknesses. We will name these thicknesses, respectively, as $d_{(+)}^{eff}$ and $d_{(-)}^{eff}$. Figs. 1a,b illustrate MS-wave helical modes in a normally magnetized ferrite disk with the "reflection walls" on planes $z = -\frac{1}{2}d_{(+)}^{eff}, \frac{1}{2}d_{(+)}^{eff}$ and $-\frac{1}{2}d_{(-)}^{eff}, \frac{1}{2}d_{(-)}^{eff}$. The pictures correspond the case when

$$p_{(+,-)} = d_{(+,-)}^{eff}.$$

Together with the interactions $\psi^{(1)} \leftrightarrow \psi^{(4)}$ and $\psi^{(2)} \leftrightarrow \psi^{(3)}$ it is interesting to consider possible interactions $\psi^{(1)} \leftrightarrow \psi^{(2)}$ and $\psi^{(3)} \leftrightarrow \psi^{(4)}$, which may appear also in the reflection points. Contrary to the continuity relations (27) and (28), for possible $\psi^{(1)} \leftrightarrow \psi^{(2)}$, $\psi^{(3)} \leftrightarrow \psi^{(4)}$ interactions one has the following inequalities in the reflection points:

$$B_r^{(1)} \neq B_r^{(2)}, \; B_\theta^{(1)} \neq B_\theta^{(2)} \quad \text{and} \quad B_r^{(3)} \neq B_r^{(4)}, \; B_\theta^{(3)} \neq B_\theta^{(4)}. \tag{29}$$

The differences between radial and azimuth components of the magnetic flux density are expressed as:

$$B_r^{(2)} - B_r^{(1)} = -\mu\left(\frac{\partial \psi^{(2)}}{\partial r} - \frac{\partial \psi^{(1)}}{\partial r}\right) + i\mu_a \tan\alpha_0\left(\frac{\partial \psi^{(2)}}{\partial \zeta} - \frac{\partial \psi^{(1)}}{\partial \zeta}\right) + i\mu_a \frac{1}{r}\left(\frac{\partial \psi^{(2)}}{\partial \phi} + \frac{\partial \psi^{(1)}}{\partial \phi}\right) \quad (30)$$

and

$$B_\theta^{(2)} - B_\theta^{(1)} = -\mu\frac{1}{r}\left(\frac{\partial \psi^{(2)}}{\partial \phi} - \frac{\partial \psi^{(1)}}{\partial \phi}\right) - \mu\tan\alpha_0\left(\frac{\partial \psi^{(2)}}{\partial \zeta} + \frac{\partial \psi^{(1)}}{\partial \zeta}\right) - i\mu_a\left(\frac{\partial \psi^{(2)}}{\partial r} + \frac{\partial \psi^{(1)}}{\partial r}\right). \quad (31)$$

We also have

$$B_r^{(4)} - B_r^{(3)} = -\mu\left(\frac{\partial \psi^{(4)}}{\partial r} - \frac{\partial \psi^{(3)}}{\partial r}\right) - i\mu_a \tan\alpha_0\left(\frac{\partial \psi^{(4)}}{\partial \zeta} - \frac{\partial \psi^{(3)}}{\partial \zeta}\right) - i\mu_a \frac{1}{r}\left(\frac{\partial \psi^{(4)}}{\partial \phi} + \frac{\partial \psi^{(3)}}{\partial \phi}\right) \quad (32)$$

and

$$B_\theta^{(4)} - B_\theta^{(3)} = -\mu\frac{1}{r}\left(\frac{\partial \psi^{(4)}}{\partial \phi} - \frac{\partial \psi^{(3)}}{\partial \phi}\right) - \mu\tan\alpha_0\left(\frac{\partial \psi^{(4)}}{\partial \zeta} + \frac{\partial \psi^{(3)}}{\partial \zeta}\right) + i\mu_a\left(\frac{\partial \psi^{(4)}}{\partial r} + \frac{\partial \psi^{(3)}}{\partial r}\right). \quad (33)$$

The right-hand sides of Eqs. (30) and (32) are real quantities while the right-hand sides of Eqs. (31) and (32) are imaginary quantities. It becomes evident that for one-dimensional arrays of helical modes such interactions in the reflection points as $\psi^{(1)} \leftrightarrow \psi^{(2)}$ and $\psi^{(3)} \leftrightarrow \psi^{(4)}$ look as nontrivial bundles which are similar to the south and north magnetic poles.

A real ferrite disk is an open thin-film structure with a small thickness/diameter ratio. In this case separation of variables is possible and one has a ferrite-rod and ferrite-slab system of equations [5,6]. For a real ferrite disk thickness $d$, one, certainly, has $\beta^{(1,4)}d \ll \beta^{(1,4)}d_{(+)}^{eff}$ and $\beta^{(2,3)}d \ll \beta^{(2,3)}d_{(-)}^{eff}$. So the virtual reflection points for helical modes are found in free space regions above and below a disk. These reflection points are, in fact, the mapping points. Figs. 2a,b illustrate the (+) and (−) resonances in a real thin-film disk. The pictures correspond, respectively, to the cases shown in Figs. 1a,b.

For MS-potential functions, in regions above and below a disk ($z \leq -\frac{1}{2}d, z \geq \frac{1}{2}d$) there are exponentially descending solutions along z axis. Following the method of separation of variables used in [5], in regions ($z \leq -\frac{1}{2}d, z \geq \frac{1}{2}d$) and for $r \leq \Re$ we describe the MS-potential function by the Bessel equation:

$$\frac{\partial^2 \psi_{(+,-)}(r)}{\partial r^2} + \frac{1}{r}\frac{\partial \psi_{(+,-)}(r)}{\partial r} + \left[\alpha_{(+,-)}^2 - \frac{\nu_{(+,-)}^2}{r^2}\right]\psi_{(+,-)}(r) = 0. \quad (34)$$

MS-potential function in the outside region $z \geq \frac{1}{2}d, r \leq \Re$ is described as:

$$\psi_{(+,-)} = (f_1)_{(+,-)} e^{-i\nu_{(+,-)}\theta} e^{-\alpha_{(+,-)}(z-\frac{1}{2}d)} \tag{35}$$

and in the outside region $z \leq -\frac{1}{2}d, r \leq \Re$ is described as:

$$\psi_{(+,-)} = (f_2)_{(+,-)} e^{-i\nu_{(+,-)}\theta} e^{\alpha_{(+,-)}(z+\frac{1}{2}d)}. \tag{36}$$

Coefficients $f_1$ and $f_2$ are amplitude coefficients.

The boundary conditions on plane surfaces of a ferrite disk are the following. For the (+) resonance there are:

$$\psi^{(1)}\Big|_{z=\left|\frac{d}{2}\right|^+} = \psi^{(1)}\Big|_{z=\left|\frac{d}{2}\right|^-}, \qquad (B_z)^{(1)}\Big|_{z=\left|\frac{d}{2}\right|^+} = (B_z)^{(1)}\Big|_{z=\left|\frac{d}{2}\right|^-}, \tag{37a}$$

$$\psi^{(4)}\Big|_{z=\left|\frac{d}{2}\right|^+} = \psi^{(4)}\Big|_{z=\left|\frac{d}{2}\right|^-}, \qquad (B_z)^{(4)}\Big|_{z=\left|\frac{d}{2}\right|^+} = (B_z)^{(4)}\Big|_{z=\left|\frac{d}{2}\right|^-}. \tag{37b}$$

For the (–) resonance there are:

$$\psi^{(2)}\Big|_{z=\left|\frac{d}{2}\right|^+} = \psi^{(2)}\Big|_{z=\left|\frac{d}{2}\right|^-}, \qquad (B_z)^{(2)}\Big|_{z=\left|\frac{d}{2}\right|^+} = (B_z)^{(2)}\Big|_{z=\left|\frac{d}{2}\right|^-}, \tag{38a}$$

$$\psi^{(3)}\Big|_{z=\left|\frac{d}{2}\right|^+} = \psi^{(3)}\Big|_{z=\left|\frac{d}{2}\right|^-}, \qquad (B_z)^{(3)}\Big|_{z=\left|\frac{d}{2}\right|^+} = (B_z)^{(3)}\Big|_{z=\left|\frac{d}{2}\right|^-}. \tag{38b}$$

We define now a four-component-function of a space z-component of a magnetic flux density. Since in a ferrite region $B_z = B_\zeta + B_\phi \sin\alpha_0$ [19], one has from Eqs. (13) – (16) after some algebraic transformations:

$$B_z^{(1,2,3,4)} = -\frac{\partial \psi^{(1,2,3,4)}}{\partial \zeta} = -\frac{\partial \psi^{(1,2,3,4)}}{\partial z}. \tag{39}$$

One can rewrite this equation as:

$$[B_z] = -\begin{pmatrix} \nabla_\| \psi^{(1)} \\ \nabla_\| \psi^{(2)} \\ \nabla_\| \psi^{(3)} \\ \nabla_\| \psi^{(4)} \end{pmatrix}. \tag{40}$$

The boundary conditions (37) and (38) can be rewritten as:

$$\psi^{(1)}\Big|_{z=\left|\frac{d}{2}\right|^+} = \psi^{(1)}\Big|_{z=\left|\frac{d}{2}\right|^-}, \qquad \left(\frac{\partial \psi^{(1)}}{\partial z}\right)_{z=\left|\frac{d}{2}\right|^+} = \left(\frac{\partial \psi^{(1)}}{\partial z}\right)_{z=\left|\frac{d}{2}\right|^-}, \qquad (41a)$$

$$\psi^{(4)}\Big|_{z=\left|\frac{d}{2}\right|^+} = \psi^{(4)}\Big|_{z=\left|\frac{d}{2}\right|^-}, \qquad \left(\frac{\partial \psi^{(4)}}{\partial z}\right)_{z=\left|\frac{d}{2}\right|^+} = \left(\frac{\partial \psi^{(4)}}{\partial z}\right)_{z=\left|\frac{d}{2}\right|^-}, \qquad (41b)$$

$$\psi^{(2)}\Big|_{z=\left|\frac{d}{2}\right|^+} = \psi^{(2)}\Big|_{z=\left|\frac{d}{2}\right|^-}, \qquad \left(\frac{\partial \psi^{(2)}}{\partial z}\right)_{z=\left|\frac{d}{2}\right|^+} = \left(\frac{\partial \psi^{(2)}}{\partial z}\right)_{z=\left|\frac{d}{2}\right|^-}, \qquad (42a)$$

$$\psi^{(3)}\Big|_{z=\left|\frac{d}{2}\right|^+} = \psi^{(3)}\Big|_{z=\left|\frac{d}{2}\right|^-}, \qquad \left(\frac{\partial \psi^{(3)}}{\partial z}\right)_{z=\left|\frac{d}{2}\right|^+} = \left(\frac{\partial \psi^{(3)}}{\partial z}\right)_{z=\left|\frac{d}{2}\right|^-}. \qquad (42b)$$

All the above analysis for magnetostatic modes in helical coordinates and symmetry breaking we made in a supposition that components $\mu$ and $\mu_a$ of the permeability tensor remain the same when one passes from a cylindrical to a helical coordinate system. It means that we suppose to have the same close-loop precession process for electrons in helical coordinates. It means, in other words, that the obtained oscillations of helical MS-wave harmonics in a ferrite disk should be considered as adiabatic variations with respect to a dynamical precession process described by the Landau-Lifshitz equation. Such angle change in the transverse component of a magnetic moment when the field about which it precesses is varied adiabatically through a closed loop was considered by Cina [21]. Because of these adiabatic variations, MS-potential eigenfunctions should, in general, pick up the Berry phase.

The properties of the (+) and (–) resonances should be characterized in terms of energy eigenstates. To solve the problem of helical-wave resonances for magnetostatic modes in a ferrite disk one should know the quantities $\nu_{(+)}$ and $\nu_{(-)}$ in Eqs. (25) and (26). These quantities cannot be found from the above analysis and are unknown a priori. For further analysis we need to make a more explicit analysis in a cylindrical coordinate system.

## 3. Circular magnetic current and gauge vector potential for magnetic-dipolar oscillating modes.

In a cylindrical coordinate system, because of separation of variables, one can formulate the eigenvalue problem for MS waves propagating in a ferrite rod [5]. Based on two MS equations:

$$\vec{H} = -\nabla \psi, \qquad \vec{B} = -\vec{\mu} \cdot \nabla \psi \qquad (43)$$

and taking into account the equation $\nabla \cdot \vec{B} = 0$, one can write the following operator equation:

$$\hat{L}V = 0, \qquad (44)$$

where

$$\hat{L} = \begin{pmatrix} (\ddot{\mu})^{-1} & \nabla \\ \nabla \cdot & 0 \end{pmatrix} \qquad (45)$$

is the differential-matrix operator and

$$V = \begin{pmatrix} \vec{B} \\ \psi \end{pmatrix} \qquad (46)$$

is the vector function included in the domain of definition of operator $\hat{L}$.

For propagating MS waves, one can represent function $V$ as

$$V = \tilde{V} e^{-i\kappa z}, \qquad (47)$$

where tilder means MS-wave membrane functions: $\tilde{V} = \begin{pmatrix} \vec{\tilde{B}} \\ \tilde{\varphi} \end{pmatrix}$, $\kappa$ is a propagation constant along $z$ axis. The eigenvalue equation for MS mode $m$ is expressed as:

$$\left(\hat{L}_\perp - i\kappa_m \hat{R}\right)\tilde{V}_m = 0, \qquad (48)$$

where

$$\hat{R} = \begin{pmatrix} 0 & \vec{e}_z \\ -\vec{e}_z & 0 \end{pmatrix}, \qquad (49)$$

subscript $\perp$ means differentiation over a waveguide cross section and $\vec{e}_z$ is a unit vector of a longitudinal $z$ axis. Integration by parts on $S$ – a square of an open MS-wave waveguide – of the integral $\int_S (\hat{L}_\perp \tilde{V})\tilde{V}^* dS$ gives the contour integral in a form $\oint_C (\tilde{B}_r \tilde{\varphi}^* - \tilde{B}_r^* \tilde{\varphi})\, dc$, where $C$ is a contour surrounding a cylindrical ferrite core and $B_r$ is a component of a membrane function of the magnetic flux density normal to contour $C$. Operator $\hat{L}_\perp$ becomes self-adjoint for homogeneous boundary conditions (continuity of $\tilde{\varphi}$ and $\tilde{B}_r$) on contour $C$.

Based on the homogeneous boundary conditions one obtains the orthogonality relation:

$$(\kappa_m - \kappa_n)\int_S \left(\hat{R}\tilde{V}_m\right)\left(\tilde{V}_n\right)^* dS = 0. \qquad (50)$$

The norm of mode $m$ is expressed as:

$$N_m = -i\omega \int_S \left(\hat{R}\tilde{V}_m\right)\left(\tilde{V}_m\right)^* dS = -i\omega \int_S \left[\tilde{\varphi}_m \left(\vec{\tilde{B}}_m\right)^* - (\tilde{\varphi}_m)^* \vec{\tilde{B}}_m\right]\cdot \vec{e}_z dS, \qquad (51)$$

where factor $i\omega$ is used as a dimensional coefficient. This norm (derived by 4) describes the average (on the period $2\pi/\omega$) power flow through a waveguide cross section.

For the quasi-monochromatic MS wave process, the energy balance equation in a waveguide section of an axially magnetized ferrite rod is written as [5]:

$$\int_{z_1}^{z_2}\int_S \nabla_\| \cdot \overline{\vec{P}}_\| \, ds \, dz = -\frac{d}{dt}\int_{z_1}^{z_2}\int_S \overline{w} \, ds \, dz, \qquad (52)$$

where $\overline{\vec{P}}_\|$ is the average (on the RF period) power flow density along a MS waveguide, $\nabla_\|$ means the longitudinal part of divergence, and $\overline{w}$ is the average density of energy. Based on this energy balance equation one can formally reduce the spectral problem to the energy eigenvalue problem [5]. Since a normally magnetized ferrite disk can be considered as a section of an axially magnetized ferrite rod, one can suppose that the energy orthogonality relation is applicable for a disk as well [5].

As we noted, formulation of the above spectral problem is based on the homogeneous boundary conditions. Continuity of the radial component of $\vec{B}$ is described as

$$\mu(H_r)_{r=\Re^-} - (H_r)_{r=\Re^+} = -i\mu_a (H_\theta)_{r=\Re^-}, \qquad (53)$$

where $H_r$ and $H_\theta$ are, respectively, radial and azimuth components of the RF magnetic field. For magnetostatic solutions: $H_r = -\frac{\partial \psi}{\partial r}$ and $H_\theta = -\frac{1}{r}\frac{\partial \psi}{\partial \theta}$. Because of a cylindrical symmetry of a ferrite rod, the membrane function $\tilde{\varphi}$ is written as $\tilde{\varphi} = \tilde{\varphi}(r)\tilde{\varphi}(\theta)$. With a supposition that an angular part is described as $\tilde{\varphi}(\theta) = e^{-i\nu\theta}$, one rewrites (53) as:

$$\mu\left(\frac{\partial \tilde{\varphi}}{\partial r}\right)_{r=\Re^-} - \left(\frac{\partial \tilde{\varphi}}{\partial r}\right)_{r=\Re^+} = -\frac{\mu_a}{\Re}\nu(\tilde{\varphi})_{r=\Re^-}. \qquad (54)$$

It becomes evident that for a given sign of $\mu_a$, the solutions for MS-wave functions depend on a sign of $\nu$. For an axially magnetized ferrite rod this fact was shown, for the first time, by Joseph and Schlömann [22]. So because of the boundary conditions we have different functions $\tilde{\varphi}$ for positive and negative directions of an angle coordinate when $0 \leq \alpha \leq 2\pi$. In other words, one can distinguish the (+) and (–) functions $\tilde{\varphi}$ – the situation described above by Eqs. (24) and (25). It means that functions $\tilde{\varphi}$ cannot be considered as the single-valued functions. We have a sequence of angular eigenvalues restricted from above and below by values equal in a modulus and different in a sign, which we denote as $\pm s^e$. The difference $2s^e$ between the largest and smallest values is an integer or zero. So $s^e$ can have values $0, \pm 1/2, \pm 1, \pm 3/2, ..., \nu/2$. At a full-angle "in-plane" rotation (at an angle equal to $2\pi$) of a system of coordinates, the "in-plane" functions $\tilde{\varphi}$ with integer values $s^e$ return to their initial states (single-valued functions) and "in-plane" functions $\tilde{\varphi}$ with the half-integer values $s^e$ will have an opposite sign (double-valued functions). The only possibility in our case is to suggest that $s^e$ are the half-integer quantities. Because of the double-valuedness properties of MS-potential functions on a lateral surface of a ferrite disk resonator, we can talk about the "spinning-type rotation" along a border contour $C$. Along with the well-known notion of the "magnetic spin" as a quantity correlated with the eigen magnetic moment of a particle, we introduce the notion of the "electric spin" as a quantity correlated with the eigen electric moment. For integer quantities $s^e$ the eigen electric moment is equal to zero, but it is non-zero for half-integer values $s^e$. One should note, however, that the "electric spin" is the boundary ("orbital") state of the magnetic system that has changed sign and thus has nothing to do with real electron spin.

The main feature of boundary condition (53) arises from the quantity of an azimuth magnetic field in the right-hand side. One can see that this is a singular field, which exists only in an infinitesimally narrow cylindrical layer abutting (from a ferrite side), to the ferrite-dielectric border. One does not have any special conditions connecting radial and azimuth components of magnetic fields on other (inner or outer) circular contours, except contour $C$. Let us introduce formally a quantity of a magnetic current:

$$\vec{j}^m(z) \equiv \frac{1}{4\pi} i\omega\mu_a \vec{H}_\theta(z). \tag{55}$$

One can rewrite the boundary condition (53) as follows:

$$\delta(r-\Re)\left[\frac{1}{4\pi}\omega\mu(H_r)_{r=\Re^-} - \frac{1}{4\pi}\omega(H_r)_{r=\Re^+}\right] = -i^m, \tag{56}$$

where $i^m$ is a density of an effective boundary magnetic current defined as:

$$i^m(z) \equiv \delta(r-\Re)\frac{1}{4\pi}i\omega\mu_a(H_\theta(z))_{r=\Re^-} = \delta(r-\Re)j^m(z). \tag{57}$$

This magnetic current cannot be considered as a single-valued function. So in the description of the MS-potential functions in a ferrite disk, taking into account the effective surface magnetic current, certain additional coordinates (additional eigenvalues and eigenfunctions) should appear on boundary contour $C$. These singular functions describing the "spin states" we will denote as $\tilde{\tilde{\varphi}}$. For $l$-th "border" eigenfunction having amplitude $f_l$, we can write: $\tilde{\tilde{\varphi}}_l = f_l e^{-iq_l\theta}$, where $q_l = l\frac{1}{2}$ and $l$ is an integer odd (positive or negative) quantity. For a certain "thickness distribution" $\tilde{\xi}(z)$ and with representation: $(H_\theta(z))_{r=\Re^-} = -A\tilde{\xi}(z)\nabla_\theta\tilde{\tilde{\varphi}}$, we have for a certain "flat" mode $\tilde{\varphi}$:

$$(i^m(z))_l = -A_l\tilde{\xi}(z)\frac{i\omega\mu_a}{4\pi\Re}\frac{\partial\tilde{\tilde{\varphi}}_l}{\partial\theta}\bigg|_{r=\Re^-} = -A_l\tilde{\xi}(z)\frac{\omega\mu_a}{4\pi\Re}q_l f_l e^{-iq_l\theta}, \tag{58}$$

where $A_l$ is an amplitude coefficient. The circular surface magnetic current does not exist due to only precession of magnetization. It appears because of the combined effect of precession in a ferrite material and "spinning rotation" caused by the special-type boundary conditions. Circulation of magnetic current along a border contour: $D_l(z) = \oint_C (i^m)_l\, dc$ gives a non-zero quantity when $q_l$ is a number divisible by $\frac{1}{2}$.

The fact that solution of our problem is dependent on a sign of $\nu$, rises a question about validity of the energy orthogonality relation [5]. For a system for which a total Hamiltonian is conserved, there should be single valuedness for egenfunctions [23]. Since the eigenstates of Eq.

(62) are not single valued, one should find a phase factor that will make the states single valued. This will give a real physical justification of existence of effective boundary magnetic current $i^m$.

The phase factor becomes evident from the following consideration. For a MS waveguide, a system of two first-order homogeneous equations considered above can be reduced to one second-order homogeneous differential equation. This is the Walker equation [24]:

$$\hat{G}\psi = 0, \qquad (59)$$

where

$$\hat{G} = -\nabla \cdot (\vec{\mu} \nabla). \qquad (60)$$

Let us represent the MS-potential wave function as

$$\psi = \tilde{\chi} \, e^{-ikz}, \qquad (61)$$

where $\tilde{\chi}$ is the MS-potential membrane function and $k$ is a propagation constant along $z$ axis. The eigenvalue equation for MS mode $q$ in an axially magnetized ferrite rod is expressed as:

$$\left(\hat{G}_\perp - k_q^2\right)\tilde{\chi}_q = 0. \qquad (62)$$

For a ferrite region we have

$$\hat{G}_\perp = \mu \nabla_\perp^2, \qquad (63)$$

where $\nabla_\perp^2$ is the two-dimensional Laplace operator. Double integration by parts on square $S$ of the integral $\int_S (\hat{G}_\perp \tilde{\chi}) \, \tilde{\chi}^* dS$ gives the following boundary conditions for self-adjointess of operator $\hat{G}_\perp$:

$$\mu \left(\frac{\partial \tilde{\chi}}{\partial r}\right)_{r=\Re^-} - \left(\frac{\partial \tilde{\chi}}{\partial r}\right)_{r=\Re^+} = 0. \qquad (64)$$

or

$$\mu (H_r)_{r=\Re^-} - (H_r)_{r=\Re^+} = 0. \qquad (65)$$

In this case the characteristic equation has a form [6]:

$$(-\mu)^{\frac{1}{2}} \left(\frac{J'_j}{J_j}\right)_{r=\Re} + \left(\frac{K'_j}{K_j}\right)_{r=\Re} = 0, \qquad (66)$$

where the Bessel function order $j$ is an integer number. It is evident that for membrane functions $\tilde{\chi}$ the (+) and (−) resonances in a normally magnetized ferrite disk are degenerate.

MS-potential functions $\tilde{\chi}$ included in the domain of definition of operator $\hat{G}_\perp$ are functions with finite energy. The boundary conditions (64) are called as the *essential boundary conditions* (EBCs). In accordance with the Ritz method it is sufficient to use basic functions from the energetic functional space with application of the essential boundary conditions [25]. The energy eigenvalue problem for MS waves in a ferrite disk resonator is formulated as the problem defined by the differential equation:

$$\hat{F}_\perp \tilde{\chi}_q = E_q \tilde{\chi}_q \tag{67}$$

together with the corresponding (essential) boundary conditions [6]. A two-dimensional ("in-plane") differential operator $\hat{F}_\perp$ and energy $E_q$ are determined as:

$$\hat{F}_\perp = \frac{1}{2} g \mu \nabla_\perp^2, \tag{68}$$

$$E_q = \frac{1}{2} g k_q^2, \tag{69}$$

where $g$ is the unit dimensional coefficient. The energy orthonormality in a ferrite disk described now as

$$(E_q - E_{q'}) \int_S \tilde{\chi}_q \tilde{\chi}_{q'}^* dS = 0 \tag{70}$$

acquires a real physical meaning. There are the Hilbert functional space of MS-potential functions $\tilde{\chi}$. Because of discrete energy eigenstates of MS-wave oscillations resulting from structural confinement in a case of a normally magnetized ferrite disk, one can consider the oscillating system as a collective motion of quasiparticles – the light magnons [6].

The energy eigenvalue problem formulated based on the EBCs shows that a ferrite disk with magnetic-dipolar-mode oscillations is a Hamiltonian system. The essential boundary conditions differ from the boundary conditions which demand continuity for normal components of $\vec{B}$. The last ones (called as *natural boundary conditions* (NBCs) [25]) are necessarily satisfied by the boundary conditions of functions $\tilde{V}$ – the functions included in the domain of definition of operator $\hat{L}_\perp$ – but not by functions with finite energy. Evidently, the equation $\nabla \cdot \vec{B} = 0$ is satisfied for the NBCs, but not for the EBCs. The eigenfunctions $\tilde{V}$ of operator $\hat{L}_\perp$ are not the single-valued functions since membrane functions $\tilde{\varphi}$ cannot be considered as single-valued ones.

To make functions $\tilde{\varphi}$ single-valued, one should find a certain phase factor. Let us formally consider two joint boundary problems: the main boundary problem and the conjugate boundary problem. The main problem is expressed by a differential equation [which is similar to Eq. (48)]:

$$\left( \hat{L}_\perp - i\beta \hat{R} \right) \tilde{V} = 0. \tag{71}$$

The conjugate problem is expressed by an equation:

$$\left( \hat{L}_\perp^\circ - i\beta^\circ \hat{R} \right) \tilde{V}^\circ = 0. \tag{72}$$

A form of differential operator $\hat{L}_\perp^\circ$ one gets from the integration by parts:

$$\int_S (\hat{L}_\perp \tilde{V})(\tilde{V}^\circ)^* dS = \int_S \tilde{V}(\hat{L}_\perp^\circ \tilde{V}^\circ)^* dS + \oint_C P(\tilde{V}, \tilde{V}^\circ) dC, \qquad (73)$$

where $P(\tilde{V}, \tilde{V}^\circ)$ is a bilinear form. For an open ferrite structure [a core ferrite region (F) is surrounded by a dielectric region (D)] the homogeneous boundary conditions for functions $\tilde{V}$ and $\tilde{V}^\circ$ give

$$\oint_C [P^{(F)}(\tilde{V}, \tilde{V}^\circ) + P^{(D)}(\tilde{V}, \tilde{V}^\circ)] dC = 0. \qquad (74)$$

In this case operator $\hat{L}_\perp$ is a self-conjugate operator. For self-conjugate operators, the orthogonality relations can be derived. When one considers functions $\tilde{V}$ and $\tilde{V}^\circ$ as the fields of modes $m$ and $n$, one obtains the orthogonality relation (50).

We demand continuity of $\tilde{\varphi}$ and $\tilde{B}_r$ on the border $C$. So the boundary condition (74) we should write as

$$\oint_C \{[(\tilde{B}_r)_{r=\Re^-} - (\tilde{B}_r)_{r=\Re^+}](\tilde{\varphi}^\circ)^*_{r=\Re} - (\tilde{\varphi})_{r=\Re}[(\tilde{B}_r^\circ)_{r=\Re^-} - (\tilde{B}_r^\circ)_{r=\Re^+}]^*\} dC = 0. \qquad (75)$$

We now uncover the expression for magnetic flux density $\vec{B}$ in Eq. (75). Since in a ferrite region $\tilde{B}_r = \mu \frac{\partial \tilde{\varphi}}{\partial r} + i\mu_a \frac{\partial \tilde{\varphi}}{\partial \alpha}$ and in a dielectric $\tilde{B}_r = \frac{\partial \tilde{\varphi}}{\partial r}$, one has

$$\oint_C [\mu \left(\frac{\partial \tilde{\varphi}}{\partial r}\right)_{r=\Re^-} - \left(\frac{\partial \tilde{\varphi}}{\partial r}\right)_{r=\Re^+}](\tilde{\varphi}^\circ)^*_{r=\Re} - (\tilde{\varphi})_{r=\Re}[\mu \left(\frac{\partial \tilde{\varphi}^\circ}{\partial r}\right)_{r=\Re^-} - \left(\frac{\partial \tilde{\varphi}^\circ}{\partial r}\right)_{r=\Re^+}]^* dC +$$
$$\oint_C [(i\mu_a \frac{\partial \tilde{\varphi}}{\partial \theta})(\tilde{\varphi}^\circ)^* - (\tilde{\varphi})(i\mu_a \frac{\partial \tilde{\varphi}^\circ}{\partial \theta})^*]_{r=\Re} dC = 0. \qquad (76)$$

In the above equation we represented a contour integral (75) as a sum of two contour integrals.

For a case of the single-valuedness, the first integral in Eq. (76) should be equal to zero [see Eq. (64)]. Since, however, functions $\tilde{\varphi}$ are not single-valued functions, the first integral in Eq. (76) is not equal to zero.

Let us introduce a new membrane function $\tilde{\eta}$:

$$\tilde{\eta}(\rho, \alpha) = \begin{cases} \gamma_+ \tilde{\varphi}_+ \\ \gamma_- \tilde{\varphi}_- \end{cases}, \qquad (77)$$

where

$$\gamma_\pm = a_\pm e^{-iq_\pm \theta}. \qquad (78)$$

The function $\tilde{\varphi}$ changes sign when $\theta$ is rotated by $2\pi$. Therefore, in order to cancel this sign change, $\gamma_{\pm}$ must change its sign to preserve the single-valued nature of $\tilde{\varphi}$. From this we conclude that $e^{-iq_{\pm}2\pi} = -1$. That is

$$q_{\pm} = l\frac{1}{2}, \tag{79}$$

where $l = \pm 1, \pm 3, \pm 5, ...$

Now we rewrite Eq. (77) as follows:

$$\tilde{\varphi}_{\pm} = \frac{1}{\gamma_{\pm}}\tilde{\eta} = \delta_{\mp}\tilde{\eta}, \tag{80}$$

where

$$\delta_{\mp} = \frac{1}{a_{\pm}}e^{-iq_{\mp}\theta} \equiv f_{\mp}e^{-iq_{\mp}\theta}. \tag{81}$$

We substitute expression (80) into Eq. (76). Since the boundary conditions for single-valued functions $\tilde{\eta}$ should correspond to the boundary conditions (64), the first integral in Eq (76) becomes equal to zero.

With use of substitution (80) we have from Eq. (76)

$$\oint_C \{[i\frac{\partial(\delta_{\mp}\tilde{\eta})}{\partial\theta}](\delta_{\mp}^{\circ}\tilde{\eta}^{\circ})^* - (\delta_{\mp}\tilde{\eta})[i\frac{\partial(\delta_{\mp}^{\circ}\tilde{\eta}^{\circ})}{\partial\theta}]^*\}_{r=\Re} dC = 0. \tag{82}$$

This gives:

$$\int_0^{2\pi} \{[\delta_{\mp}(\delta_{\mp}^{\circ})^*][(i\frac{\partial\tilde{\eta}}{\partial\theta})(\tilde{\eta}^{\circ})^* - (\tilde{\eta})(i\frac{\partial\tilde{\eta}^{\circ}}{\partial\theta})^*]\}_{r=\Re} d\theta +$$
$$\int_0^{2\pi} \{[\tilde{\eta}(\tilde{\eta}^{\circ})^*][(i\frac{\partial\delta_{\mp}}{\partial\theta})(\delta_{\mp}^{\circ})^* - (\delta_{\mp})(i\frac{\partial\delta_{\mp}^{\circ}}{\partial\theta})^*]\}_{r=\Re} d\theta = 0. \tag{83}$$

Both functions, $\delta_{\mp}$ and $\eta$, describe a periodic process with respect to angle $\theta$. Let us introduce a generalized periodic function $y$ and consider the eigenvalue equation

$$i\frac{\partial}{\partial\theta}y = uy, \tag{84}$$

where $u$ is a real quantity. We introduce now a problem with eigenvalue equation conjugate to Eq. (84) (with eigen function $y^{\circ}$ and eigenvalue $u^{\circ}$) and consider an integral $\int_0^{2\pi}(i\frac{\partial y}{\partial\theta})(y^{\circ})^* d\theta$.

Using integration by parts of this integral, one finds that when $u$ (and $u^{\circ}$) are integer numbers (including 0): $u = 0, \pm 1, \pm 2, \pm 3, ...$ and when $u$ (and $u^{\circ}$) are half-integer numbers:

$u = \pm\frac{1}{2}, \pm\frac{3}{2}, \pm\frac{5}{2}, \ldots$, just only in these two separate cases operator $\hat{J}_z \equiv i\frac{\partial}{\partial \theta}$ is a self-conjugate operator and one can write the orthogonality relation:

$$(u - u°)\int_0^{2\pi} y(y°)^* d\theta = 0. \tag{85}$$

For any mixed situation (when, for example, $u$ is an integer number and $u°$ is a half-integer number), functions $y$ and $y°$ are not mutually orthogonal and, therefore, integral in Eq. (96) is not equal to zero. In other words, the spectral problems for integer egenvalues $u$ should be considered separately from the spectral problem for half-integer eigenvalues $u$. Based on this consideration of the orthogonality relation for generalized functions $y$, one should conclude that $\delta_\mp(\delta_\mp°)^* = 1$ in the first integral of Eq. (98), while $\tilde{\eta}(\tilde{\eta}°)^* = 1$ in the second integral of Eq. (83).

The first integral in Eq. (97) is evidently equal to zero. So, as a result, one has from Eq. (83):

$$\int_0^{2\pi} \{[(i\frac{\partial \delta_\mp}{\partial \theta})(\delta_\mp°)^* - (\delta_\mp)(i\frac{\partial \delta_\mp°}{\partial \theta})^*]\}_{r=\Re} d\theta = 0. \tag{86}$$

The transformation (77) restores the single valuedness, but now there is a nonzero vector-potential-type term:

$$a_\theta \equiv \int_0^{2\pi} [(i\frac{\partial \delta_\mp}{\partial \theta})(\delta_\mp°)^*]_{r=\Re} d\theta = -l\frac{1}{2}. \tag{87}$$

Since $l = \pm 1, \pm 3, \pm 5, \ldots$, there are the positive and negative vector-potential-type terms.

Function $\tilde{\eta}$ is identical to function $\tilde{\chi}$. The confinement effect for magnetic-dipolar oscillations requires proper phase relationships to guarantee single-valuedness of the wave functions. To compensate for sign ambiguities and thus to make wave functions single valued we added a vector-potential-type term to the MS-potential Hamiltonian. This procedure is similar to the procedure made by Mead [26] for the Born-Oppenheimer wave functions. The corresponding flux of pseudo-electric field $\vec{e}$ (the gauge field) through a circle of radius $\Re$ is obtained analogously to [26]:

$$\int_S \vec{e} \cdot d\vec{S} = \oint_C \vec{a}_\theta \cdot d\vec{C} = \Phi^e, \tag{88}$$

where $\Phi^e$ is the flux of pseudo-electric field. The energy levels are periodic in the electric flux $\Phi^e$. There should be the positive and negative fluxes. These different-sign fluxes should be inequivalent to avoid the cancellation.

## 4. Helical MS modes as a topological effect

We found that restoration of the single valuedness of the MS-potential eigenstates leads to nonzero vector potentials. This is a topological vector potential which shows the presence of the Berry phase. The Berry-phase curvature is adjusted to restore single-valuedness. What is most characteristic for the concept of Berry's phase is the existence of a continuous parameter space in which the state of the system can travel on a closed loop.

The "spinning" states considered above are incorporated into the boundary conditions. The closed-loop traveling in our case means simultaneous rotation and longitudinal motion on a lateral surface of a ferrite disk. The line is a helix on a cylinder of radius $\Re$. The helix on a circular cylinder is a geodesic line. To get a closed loop one should connect two types of helices – the right-hand and the left-hand helices – in the virtual reflection points on planes $z = -\frac{1}{2}d_{(+,-)}^{eff}, \frac{1}{2}d_{(+,-)}^{eff}$.

The Berry phase gives a general evidence for the existence of a topological phase for a cyclic Hamiltonian system. For the Hamiltonian system described by Eq. (67), the helical motion analyzed above in Section 2 of the paper could be considered as the adiabatic process. In this case we should suppose that the ground energy eigenstate is well separated from any other state to which transitions may be simulated by the helical motion. We have the coupling between the rotating-term magnetization and the helical-line way. In his work [21], Cina discussed the question of a geometric phase in spin precession. As a result, one has the geometrical phase factors of spin systems in which the magnetic field undergoes an adiabatic excursion. The gauge potentials can be generalized to systems where the slowly varying parameters are no longer external, but are themselves quantized. In a case of molecular physics, these are the nuclear coordinates [27]. In our case of magnetic-dipolar modes, these are the "spinning" boundary coordinates on a lateral coordinates of a normally magnetized ferrite disk. Magnetic fields with the $\phi$- and $\zeta$-helical components are varied adiabatically through a closed loop. In a helical coordinate system, a transverse component of a magnetic moment $\vec{m}$ precesses about the field with the $\zeta$-component. We illustrate in Fig.3 the RF magnetization evolution in a ferrite disk with virtual reflection planes for a case of the (+) resonance.

Spin precession of electrons in a cyclic motion can lead to various interference phenomena such as an oscillating persistent current. This gives a physical justification of a surface magnetic current $i^m$ which we formally introduced above [see Eqs. (57), (58)]. Non-zero circulation of current $i^m$ gives the eigen electric moment – the anapole moment [11]. The existence of such a moment was experimentally shown in [7].

The MS-mode spectral problem obtained based on the essential boundary conditions give energy eigenstes. These eigenstates are characterized by integer azimuth numbers and can be considered as certain "orbit" states. The helical-wave motion should be described in terms of eigenstates as the adiabatic evolution of a general state.

Suppose that we have an azimuth number $j$ in Eq. (66) equal to unit. For the (+) and (–) resonances (corresponding to the interactions $\psi^{(1)} \leftrightarrow \psi^{(4)}$ and $\psi^{(2)} \leftrightarrow \psi^{(3)}$), a total period should correspond to the $4\pi$ rotation. In other words, after the "orbit-state" $2\pi$ rotation the (+) [or (–)] resonant mode should acquire an opposite phase. This gives us a possibility to write for azimuth numbers in Eqs. (25), (26):

$$\nu_{(+,-)} = \pm(w - \bar{p}\beta) = \pm\left(\frac{1}{2} - 1\right) = \mp\frac{1}{2}. \tag{89}$$

The above analysis of the origin of the geometrical phase of precessing spins in a ferrite-disk system with a smooth close-loop time excursion of the magnetic field could be also extended to the momentum space. Following Eq. (69), one sees that the light-magnon energy is proportional to a squared wave number. One can consider Eq. (69) as an equation which describes an unperturbed motion of a "free" light magnon propagating along $z$ axis. Suppose now that we impose on this motion process a helical "modulation" with wave numbers $\beta^{(i)} \equiv \beta^{(1,4)}$ and $w^{(i)} \equiv w^{(1,4)}$ or a helical "modulation" with wave numbers $\beta^{(j)} \equiv \beta^{(2,3)}$ and $w^{(j)} \equiv w^{(2,3)}$ [see Eqs. (21), (22)].

As it was discussed by Waldron [19], there are two types of helical spaces: the simply connected helical space and the multiply connected helical space. In a simply connected space, the helical surfaces $\zeta = 0$ and $\zeta = p$ constitute barriers which cannot be crossed. A point can be expressed in one way only, by suitable values of $r$ and $\phi$, and a value of $\zeta$ lying between 0 and $p$. By crossing the surfaces $\zeta = 0$ and $\zeta = p$, one goes outside the system. In a multiply connected space, the helical surfaces $\zeta = 0$ and $\zeta = p$ are not barriers. When we cross the surfaces $\zeta = 0$ and $\zeta = p$, we move to another part of the system. Any point can be represented in an infinite number of ways by a unique value of $r$ and a value of $\zeta$, which depends on the value assigned to the system. For any point $A$ with coordinates $(r, \phi, \zeta)$, the point $B$, being distant with a period of a helix, is characterized by coordinates $(r, \phi, \zeta + p)$ or by coordinates $(r, \phi + 2\pi, \zeta)$. The regions between the surfaces $\zeta = np$ and $\zeta = (n+1)p$, for all integer numbers $n$, are continuous in a multiply connected space.

In our case we have the multiply connected helical space. For the (+) resonance, running helical modes with $\beta^{(i)}$ and $w^{(i)}$ are the reciprocal modes in a multiple connected helical space. The helical modes with $\beta^{(j)}$ and $w^{(j)}$ are another-type reciprocal modes in a multiple connected helical space. We consider wave vectors in a helical space, which are directed along a helix [19]:

$$\vec{\tau} = (\tau_\zeta, \tau_\phi) \equiv \left( \beta, \frac{r}{\cos \alpha_0} w \right). \tag{90}$$

Similar to a one-dimensional-crystal model of the reciprocal lattice [28], we chose the function $\tilde{\chi}_{q,\beta^{(i,j)}}$ to be periodic in $\beta^{(i)}$ or $\beta^{(j)}$ with the periods $2\pi / d^{eff}_{(+,-)}$.

The phase run in Eq. (87) is due to a topological effect. One has a vector potential (87) dependent on a cyclic parameter. Since there are no boundaries for the cyclic motion, the integral (87) becomes finite if the vector potential has nontrivial topology. The cyclic motion behaves exactly like traveling in a periodic structure. We can consider helical modes with periodic conditions along $z$ axis as curvature in the momentum space made out of the Bloch wave function. The vector potential in $\vec{\tau}$-space is the gauge field which is defined in terms of the Bloch state $|q(\vec{\tau}^{(i,j)})\rangle$ as

$$a_{q\beta}(\vec{\tau}^{(i,j)}) = i \left\langle q(\vec{\tau}^{(i,j)}) \left| \frac{\partial}{\partial \beta^{(i,j)}} \right| q(\vec{\tau}^{(i,j)}) \right\rangle. \tag{91}$$

The gauge change is compensated by the fact that

$$\tilde{\chi}_{q,\beta^{(i,j)} + \frac{2\pi}{d^{eff}_{(+,-)}}}(z) = e^{-i \frac{2\pi}{d^{eff}_{(+,-)}} z} \tilde{\chi}_{q,\beta^{(i,j)}}(z). \tag{92}$$

We have also:

$$|\tilde{\chi}(d^{eff}_{(+,-)})\rangle = e^{-i(\beta^{(i,j)} d^{eff}_{(+,-)})} |\chi(0)\rangle. \tag{93}$$

Because of the special helical topology of our "Brillouin zone" $[-\pi/d_{(+,-)}^{eff}, \pi/d_{(+,-)}^{eff}]$, a nonzero Berry phase is shown to exist in a one-dimensional parameter space. There are two parameters (two parameter spaces): $\vec{\beta}^{(i)}$ and $\vec{\beta}^{(j)}$. For these two parameter spaces one has from Eq. (67):

$$\hat{F}_\perp(\vec{\beta}^{(i)})\left|\tilde{\chi}(\vec{\beta}^{(i)})\right\rangle = E(\vec{\beta}^{(i)})\left|\tilde{\chi}(\vec{\beta}^{(i)})\right\rangle \tag{94}$$

and

$$\hat{F}_\perp(\vec{\beta}^{(j)})\left|\tilde{\chi}(\vec{\beta}^{(j)})\right\rangle = E(\vec{\beta}^{(j)})\left|\tilde{\chi}(\vec{\beta}^{(j)})\right\rangle. \tag{95}$$

The Berry's phases are defined, respectively as

$$\gamma^{(i)} = i \int_{-\pi/d_{(+)}^{eff}}^{\pi/d_{(+)}^{eff}} \left\langle \tilde{\chi} \middle| \nabla_{\beta^{(i)}} \tilde{\chi} \right\rangle \cdot d\vec{\beta}^{(i)} \tag{96}$$

and

$$\gamma^{(j)} = i \int_{-\pi/d_{(-)}^{eff}}^{\pi/d_{(-)}^{eff}} \left\langle \tilde{\chi} \middle| \nabla_{\beta^{(j)}} \tilde{\chi} \right\rangle \cdot d\vec{\beta}^{(j)}. \tag{97}$$

The fact that our "Brillouin zone" is a helix gives us a possibility to see that even in a one-dimensional parameter space a nonvanishing Berry phase can appear. This resembles the case of special torus topology in periodic solids [28]. In further publication we suppose to analyze a closed orbit in a full wave-vector helical space $\vec{\tau} = (\tau_\zeta, \tau_\phi)$.

## 5. Conclusion

The long-range magnetic-dipolar field variations have no influence on the character of local spin precession. In a case of magnetic-dipolar oscillations in a normally magnetized ferrite disk, MS-potential wave functions may acquire a special physical meaning: the quantum-like properties described by the Schrödinger-like equation. Together with such similarity with semiconductor quantum dots as discrete energy levels due to confinement phenomena, magnetic-dipolar ferrite particles show other, very unique, properties attributed to the quantized-like systems. There are special symmetry characterizations of the magnetic-dipolar oscillations. Topological phase curvature for magnetic-dipolar modes appears explicitly due to the Landau-Lifshitz motion equation taking into account handedness properties. For solutions of the LL equation one has different signs of the off-diagonal components of the permeability tensor in the right-handed and left-handed coordinate systems.

An analysis shows that there are four magnetic-dipolar helical waves in a ferrite disk. One can distinguish two pairs of these waves. In every pair the waves are mutually reciprocal. For mutually reciprocal waves one can consider the closed-loop way of propagation leading to appearance of resonances. These resonances are characterized by different directions of azimuth rotation.

The properties of the helical resonances should be correlated with an analysis of energy eigenstates. For a system for which a total Hamiltonian is conserved, there should be single valuedness for egenfunctions. Because of the boundary problem, the eigenstates of magnetic-dipolar modes are not single valued. To make the states single valued we found a certain phase factor. We showed that dynamical symmetry breaking leads to appearance of circular magnetic currents and eigen electric moments – the anapole moments – for magnetic-dipolar-wave processes in normally magnetized thin-film ferrite disks.

Restoration of the single valuedness of the MS-potential eigenstates leads to nonzero vector potentials. This is a topological vector potential which shows the presence of the Berry phase. We showed that an analysis of the origin of the geometrical phase of precessing spins in a ferrite-disk system with a smooth close-loop time excursion of the magnetic field can be also extended to the momentum space.

**Figure captions**

<u>Fig. 1</u>
(a) The (+) resonance caused by the $\psi^{(1)} \leftrightarrow \psi^{(4)}$ interaction (arrows show directions of propagation for helical MS modes).
(b) The (–) resonance caused by the $\psi^{(2)} \leftrightarrow \psi^{(3)}$ interaction (arrows show directions of propagation for helical MS modes).

<u>Fig. 2</u>
(a) The (+) resonance ($\psi^{(1)} \leftrightarrow \psi^{(4)}$ interaction) in a real ferrite disk
(b) The (–) resonance ($\psi^{(2)} \leftrightarrow \psi^{(3)}$ interaction) in a real ferrite disk

<u>Fig. 3</u>. RF magnetization evolution for the (+) resonance

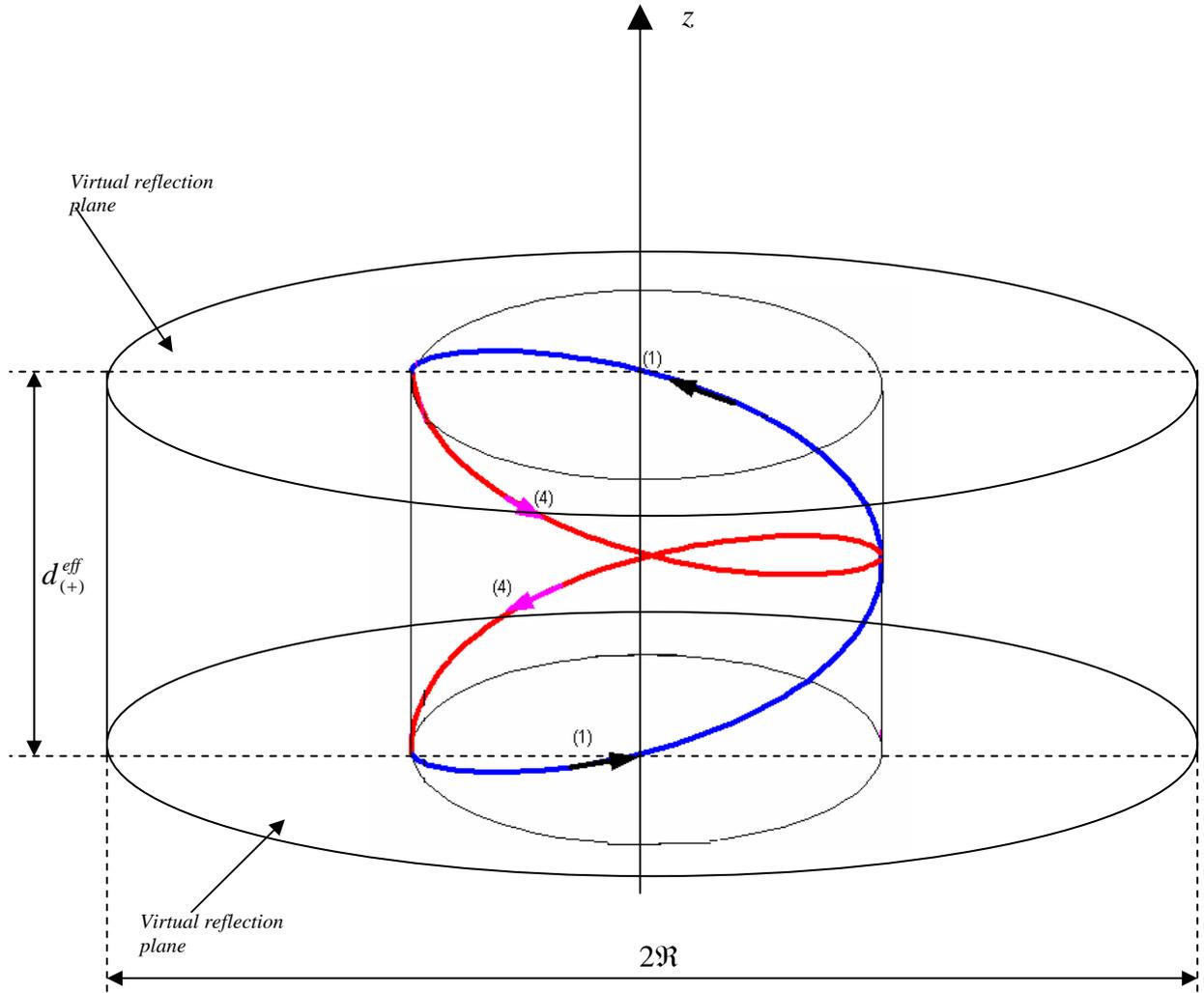

<u>Fig 1a</u>: The (+) resonance caused by the $\psi^{(1)} \leftrightarrow \psi^{(4)}$ interaction (arrows show directions of propagation for helical MS modes)

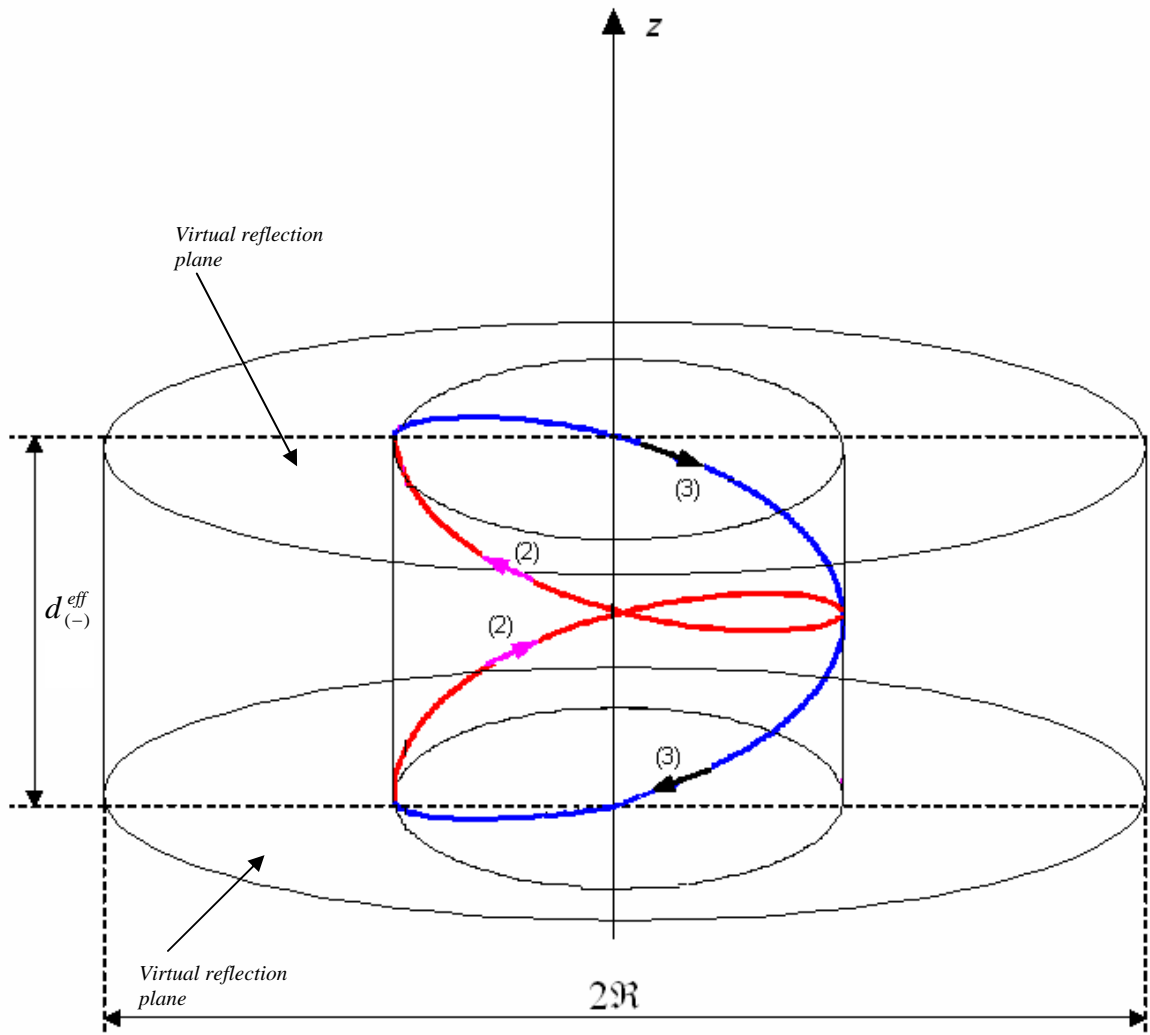

<u>Fig 1b</u>: The (–) resonance caused by the $\psi^{(2)} \leftrightarrow \psi^{(3)}$ interaction (arrows show directions of propagation for helical MS modes)

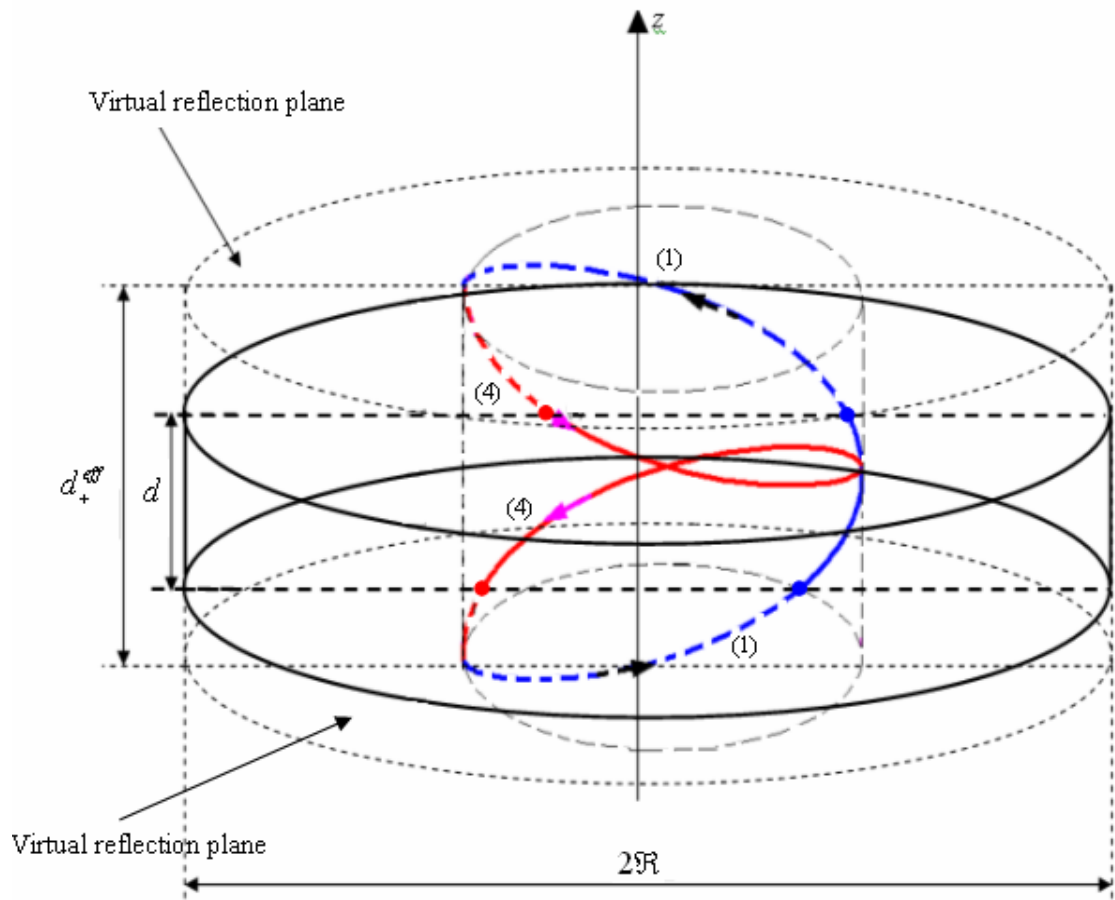

<u>Fig. 2a:</u>  The (+) resonance ($\psi^{(1)} \leftrightarrow \psi^{(4)}$ interaction) in a real ferrite disk

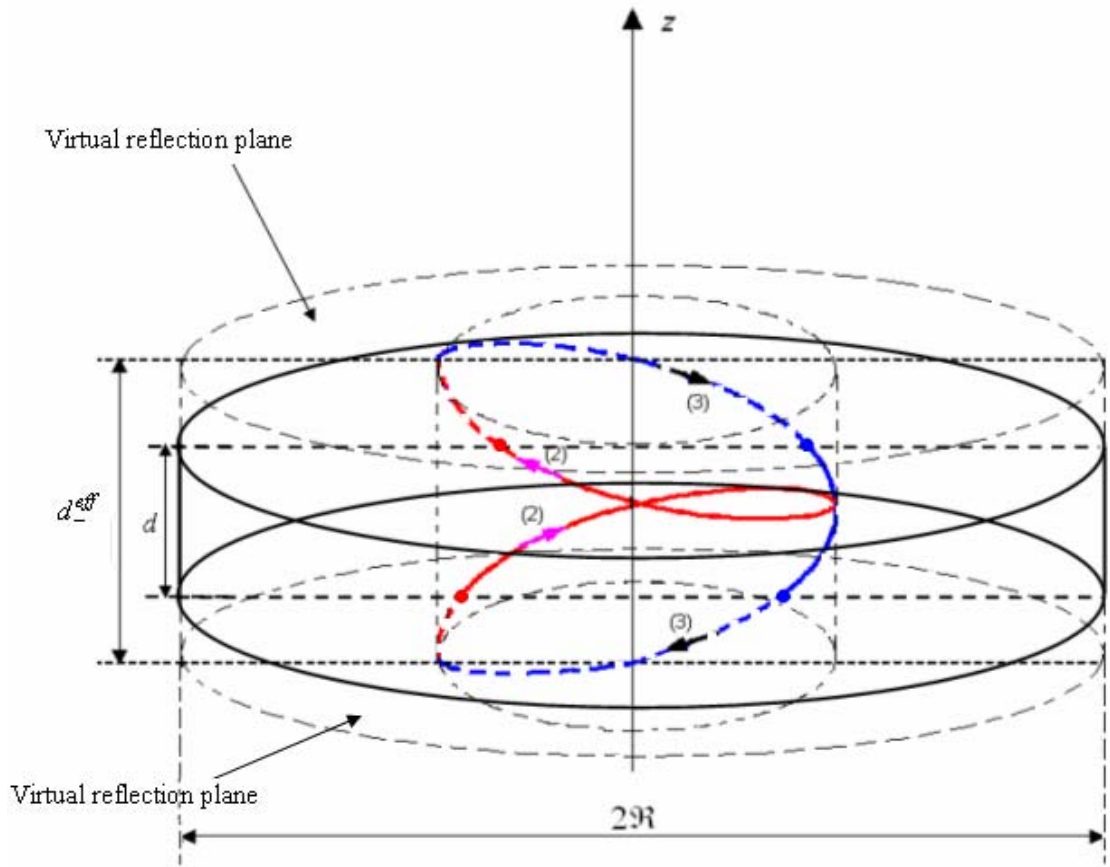

<u>Fig. 2b:</u> The (–) resonance ($\psi^{(2)} \leftrightarrow \psi^{(3)}$ interaction) in a real ferrite disk

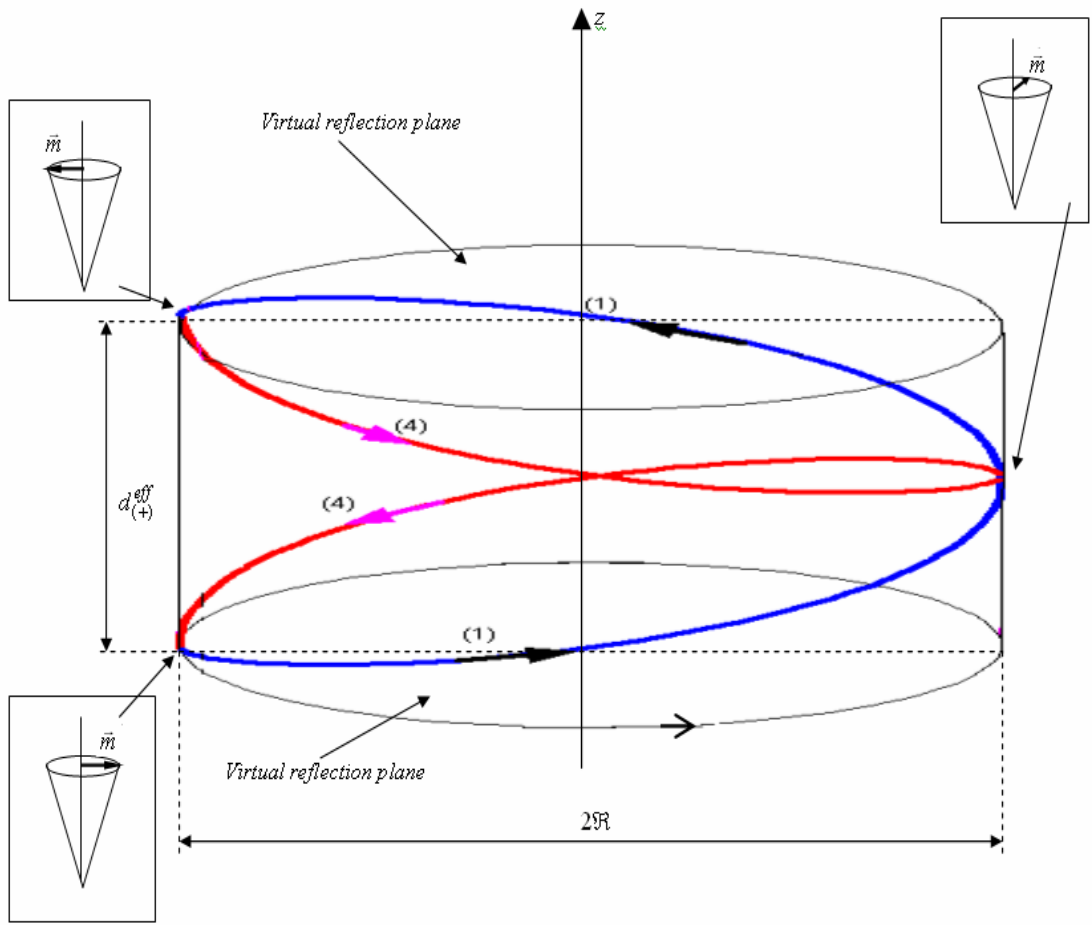

Fig. 3. RF magnetization evolution for the (+) resonance